\documentclass[aps, prd, twocolumn, notitlepage, nofootinbib]{revtex4-1}
\usepackage{graphicx}
\usepackage{float}
\usepackage{amsmath}
\usepackage{color}
\usepackage{tensor}
\usepackage{epstopdf}
\usepackage{xcolor}
\usepackage{wasysym}
\usepackage{hyperref}
\hypersetup{
    colorlinks,
    linkcolor={blue!50!black},
    citecolor={red!50!black},
    urlcolor={blue!80!black}
}

\usepackage{multibib}

\newcommand{\be}{\begin{equation}}
\newcommand{\ee}{\end{equation}}
\newcommand{\ba}{\begin{eqnarray}}
\newcommand{\ea}{\end{eqnarray}}

\newcommand{\beq}{\begin{equation}}
\newcommand{\eeq}{\end{equation}}
\newcommand{\beqa}{\begin{eqnarray}}
\newcommand{\eeqa}{\end{eqnarray}}
\newcommand{\nn}{\nonumber}

\begin{document}

\title{Shadows, Signals, and Stability in Einsteinian Cubic Gravity}

\author{Robie A. Hennigar}
\email{rhennigar@uwaterloo.ca}
\affiliation{Department of Physics and Astronomy, University of Waterloo, Waterloo, Ontario, Canada, N2L 3G1}

\author{Mohammad Bagher Jahani Poshteh}
\email{mb.jahani@iasbs.ac.ir }
\affiliation{Department of Physics and Astronomy, University of Waterloo, Waterloo, Ontario, Canada, N2L 3G1}
\affiliation{Department of Physics, Institute for Advanced Studies in Basic Sciences (IASBS), Zanjan 45137-66731, Iran}

\author{Robert B. Mann}
\email{rbmann@uwaterloo.ca}
\affiliation{Department of Physics and Astronomy, University of Waterloo, Waterloo, Ontario, Canada, N2L 3G1}
\affiliation{Perimeter Institute, 31 Caroline St. N., Waterloo,
Ontario, N2L 2Y5, Canada}

%


\begin{abstract}
We conduct a preliminary investigation into the phenomenological implications of 
Einsteinian cubic gravity (ECG), a 4-dimensional theory of gravity cubic in curvature 
of interest for its unique formulation and properties.  We find an analytic approximation
for a spherically symmetric black hole solution to this theory using a  continued fraction
ansatz.  This  approximate solution is valid everywhere outside of the horizon and  we use it  to study the orbit of massive test bodies near a black hole, specifically computing the innermost stable circular orbit.    We compute constraints on
the ECG coupling parameter  imposed by Shapiro time delay.  We then compute the shadow of
an ECG black hole and find it to be larger than its Einsteinian counterpart in general relativity
for the same value of the mass.  Applying   our results to Sgr A*, we find that departures
from general relativity are small but in principle distinguishable.
\end{abstract}

\maketitle

\section{Introduction}

The study of higher curvature corrections to general relativity has attracted significant attention in recent decades, as these corrections seem to be generic consequences of quantizing gravity. Higher curvature corrections can lead to a renormalizable theory of quantum gravity~\cite{Stelle}, and it is also understood that low energy effective actions derived from string theory result in various higher-derivative gravities~\cite{Zwiebach,Metsaev,Myers87}.  More broadly, higher curvature theories have been exploited as toy models within  the AdS/CFT correspondence, to make contact with a wider class of CFTs and study effects beyond the large $N$ limit~\cite{Kats,Myers10,Hofman,Myers11,Myers08}.

A problem plaguing most higher curvature theories is
that the linearized equations of motion allow for negative energy excitations, or ghost-like particles~\cite{Deser}. Therefore, attention has been devoted to present theories of gravity which have no ghost degrees of freedom in their propagator~\cite{Modesto,Biswas1,Biswas2}. The most general class of theories that are ghost-free on any background is Lovelock gravity, which is the natural generalization of Einstein gravity to higher dimensions~\cite{Lovelock:1971yv}. However, Lovelock terms of order $k$ in the curvature are topological in $d = 2k$ dimensions, and vanish identically for $d < 2k$.  Quasi-topological theories~\cite{Myers:2010ru, Oliva:2010eb, Cisterna:2017umf} provide additional examples in five and higher dimensions and have led to a number of interesting results in the context of holography~\cite{Myers:2010jv, Myers:2010xs}.  However, all quasi-topological theories are trivial in four dimensions.

Recently, a new theory of gravity has been proposed in~\cite{Pablo1} which, in four dimensions, is the most general up-to-cubic-order-in-curvature dimension-independent theory of gravity that shares its graviton spectrum with Einstein theory on constant curvature backgrounds. This theory, which is coined {\it Einsteinian cubic gravity} (ECG), is neither topological nor trivial in four dimensions.  ECG belongs to a class of theories that generalizes the quasi-topological theories --- \textit{generalized quasi-topological gravity}~\cite{Hennigar:2017ego, Ahmed:2017jod} --- and black hole solutions in these theories have been shown to have a number of interesting properties.

The ECG field equations admit natural generalizations of the Schwarzshild solution, i.e. static, spherically symmetric (SSS) solutions with a single metric function,
\beq
ds^2=-fdt^2+\frac{dr^2}{f}+r^2d\Omega^2_{(2)} \, .
\eeq
This non-trivial fact not only significantly simplifies the study of black holes in the theory, but is also responsible for the absence of ghosts and integrability: the field equation is a total derivative, and gives a non-linear second order differential equation upon integration, with the integration constant being related to the mass~\cite{Bueno:2017sui}.  This integrability allows for exact, analytic studies of   black hole thermodynamics, despite the lack of exact solutions\footnote{Note that, in the special case of the critical limit of the theory, certain exact solutions can be found~\cite{Feng:2017tev}.}  to the field equations~\cite{Robie17,Bueno:2016lrh}. These studies have revealed that small, asymptotically flat black hole solutions become stable, a result that may have implications in light of the information loss problem~\cite{Bueno:2017qce}. Studies of the thermodynamics of AdS black branes have revealed novel phase structure, suggesting this class of theories will provide rich holographic toy models~\cite{Hennigar:2017umz}. 

One of the most interesting aspects of ECG is that it is non-trivial in four dimensions. Consequently it is of phenomenological interest, all the more so because no dimensional reduction is required to interpret its solutions. The aim of this paper is to begin an investigation of how compatible ECG is with observational tests. Of particular interest will be constraints that arise from solar system tests and potential signatures from black hole shadows that could be constrained by the Event Horizon Telescope (EHT)~\cite{eht}. The latter will provide important constrains on any deviations from general relativity in the strong field limit.

An obstacle to performing these studies is the lack of an analytic solution, combined with the difficulty of producing a numerical one. To remedy this, we employ a continued fraction ansatz and obtain a highly accurate analytic approximate solution to the field equations. Similar techniques have recently been applied with success in a variety of contexts~\cite{Rezzolla:2014mua, Kokkotas:2017zwt, Kokkotas:2017ymc, Konoplya:2016jvv}. The continued fraction approximation is not only useful for the present analysis, but will prove useful in future investigations, e.g. concerning quasi-normal modes.

With the continued fraction solution in hand, we study new properties of the SSS black hole in ECG. Specifically we analyze the motion of particles around the black hole, constraining the coupling with solar system tests, and investigating the properties of its shadow~\cite{synge, Bardeen, Vries, Bambi1,Eiroa1,Eiroa2,Eiroa3,Bambi2} (see also~\cite{Grenzebach}). 

We constrain the  ECG coupling constant $\lambda$ using Shapiro time delay, which is the tightest constraint provided by solar system tests. We find that ECG can be compatible with solar system tests whilst maintaining relatively large values of the coupling. Furthermore, we find that the radius of the innermost stable circular orbit (ISCO) around an SSS ECG black hole and the angular momentum of a test body at this radius increase with increasing $\lambda$ as compared to their corresponding values in general relativity.

We likewise employ the continued fraction metric function to study   null  geodesic   around ECG black holes. We find in general that its shadow is  enlarged compared to 
a non-rotating black hole in general relativity.  We apply our results to the supermassive black hole Sagittarius A* (Sgr A*) at the center of our Galaxy, and show that the angular radius of the shadow  increases with increasing $\lambda$   by an amount tantalizingly close to what could be experimentally detected, whilst maintaining consistency with solar system tests.  This suggests that EHT observations could (at least in principle) provide important  constraints on $\lambda$ and on ECG in general.

The outline of out paper is as follows. In the next section we review the near horizon and asymptotic solution as well as the numeric one. Then, by using a continued fraction expansion we obtain the approximate analytic solution. In Sec.~\ref{sec:BH_props} we study some of the properties of the black holes in ECG and investigate the orbit of massive particles around it. In Sec.~\ref{sec:testECG} we constrain the coupling constant of ECG by using Shapiro test in solar system. In Sec.~\ref{sec:shadow} we study the null geodesics in ECG and present our results for Sgr A* shadow in Sect.~\ref{sec:sgr}. We conclude our paper in~\ref{sec:con}. A number of useful results are summarized in the appendices. We work in units where $G = c= 1$.

\section{Black Holes in Einsteinian Cubic Gravity}

The action for ECG reads,
\beq 
S = \frac{1}{16 \pi} \int d^4 x \sqrt{-g} \left[R - \frac{\lambda}{6} \mathcal{P}\right],
\eeq
where $R$ is the usual Ricci scalar and
\begin{align}
\mathcal{P} =& \, 12 R_a{}^b{}_c{}^d R_b{}^e{}_d{}^f R_e{}^a{}_f{}^c + R_{ab}^{cd}R_{cd}^{ef}R_{ef}^{ab} 
\nn\\
&- 12 R_{abcd}R^{ac}R^{bd} + 8 R_a^b R_b^c R_c^a \, .
\end{align} 
We restrict ourselves to asymptotically flat, static and spherically symmetric vacuum black holes.  In this case, the only independent field equation is,
\begin{align}\label{eqn:feq}
&-(f-1)r - \lambda \bigg[\frac{f'^3}{3} + \frac{1}{r} f'^2 - \frac{2}{r^2} f(f-1) f' 
\nn\\
&- \frac{1}{r} f f'' (rf' - 2(f-1)) \bigg] = 2 M,
\end{align} 
where a prime denotes differentiation with respect to $r$.  The quantity $M$ appearing on the right-hand side of the equation is the ADM mass of the black hole~\cite{Bueno:2016lrh, Hennigar:2017ego}, and we will assume $\lambda > 0$ in what follows.  

Unfortunately, the field equations cannot be solved analytically (except in certain special cases~\cite{Feng:2017tev}), and either numerical or approximate solutions (or some combination) must be computed to make progress.  We will review the construction of a numerical solution, before presenting a continued fraction expansion that provides an accurate and convenient approximation of the solution everywhere outside of the horizon.  

We begin by solving the field equations via a series expansion near the horizon using the   ansatz  
\be\label{eqn:nh_expand} 
f_{\rm nh}(r) = 4 \pi T (r-r_+) + \sum_{n=2}^{n = \infty} a_n (r-r_+)^n \, ,
\ee 
which ensures that the metric function vanishes linearly at the horizon ($r=r_+$), and $T = f'(r_+)/4\pi$ is the Hawking temperature.  Substituting this ansatz into the field equations~\eqref{eqn:feq} allows one to solve for the temperature and mass in terms of $r_+$ and the coupling $\lambda$:
\begin{align}\label{eqn:mass_temp}
M &= \frac{r_+^3}{12 \lambda^2} \left[r_+^6 + (2 \lambda - r_+^4) \sqrt{r_+^4 + 4 \lambda} \right] \, ,
\nn\\
T &= \frac{r_+}{8 \pi \lambda} \left[ \sqrt{r_+^4 + 4 \lambda} - r_+^2 \right] \, .
\end{align}   
One then finds that $a_2$ is left undetermined by the field equations, while all $a_n$ for $n > 2$ are determined by messy expressions involving $T$, $M$, $r_+$, and $a_2$.  

We now consider an expansion of the solution in the large-$r$ asymptotic region. To obtain this, we linearize the field equations about the Schwarzschild background:
\be 
f_{\rm asymp} = 1 - \frac{2 M}{r} + \epsilon h(r),
\ee
where $h(r)$ is to be determined by the field equations, and we linearize the differential equation by keeping terms only to order $\epsilon$, before setting $\epsilon = 1$.  The resulting differential equation for $h(r)$ takes the form
\be 
h'' + \gamma(r) h' - \omega(r)^2 h = g(r),
\ee
where
\begin{align}
\gamma(r) &= - \frac{2(M-r)}{(2M - r) r} \, , 
\nn\\
\omega^2(r) &=  \frac{r^6 + 56 M^2 \lambda -12 M r \lambda}{6 M r^2(r - 2M) \lambda} \, ,
\nn\\
g(r) &= - \frac{2M(46 M - 27 r)}{9(2M-r)r^3} \, .
\end{align}
In the large $r$ limit, the homogenous equation reads
\be 
h_h'' - \frac{2}{r} h_h - \frac{r^3}{6 M \lambda }h_h = 0 \, ,
\ee
and can be solved exactly in terms of Bessel functions:
\begin{align} 
h_h &= r^{3/2} \bigg[\tilde{A} I_{-\frac{3}{5}}\left(\frac{2 r^{5/2}}{5 \sqrt{ 6M \lambda }}\right) + \tilde{B}  K_{\frac{3}{5}}\left(\frac{ 2 r^{5/2}}{5 \sqrt{6 M \lambda }}\right)\bigg],
\end{align}
where $I_\nu(x)$ and $K_\nu (x)$ are the modified Bessel functions of the first and second kinds, respectively. To leading order in large $r$, this can be expanded as
\be\label{eqn:asymp_homog} 
h_h(r) \approx A r^{1/4} \exp\left[  \frac{2 r^{5/2}}{5 \sqrt{6 M\lambda} }\right] + B r^{1/4} \exp\left[ \frac{ -2 r^{5/2}}{5 \sqrt{6 M\lambda} }\right],
\ee
where we have absorbed various constants into the definitions of $A$ and $B$ (compared to $\tilde{A}$ and $\tilde{B}$).  Thus, the homogenous solution consists of a growing mode and a decaying mode.  Asymptotic flatness demands that we set $A = 0$, while the second term decays super-exponentially and can therefore be neglected.\footnote{This assumes that $\lambda > 0$.  In cases where $\lambda < 0$, the homogeneous solution contains oscillating terms that spoil the asymptotic flatness.  The only viable solution in this case is to set the homogenous solution to zero.}

More relevant is the particular solution, which reads
\begin{align}
h_{\rm p} &= -\frac{36 \lambda M^2 }{r^6} + \frac{184}{3} \frac{\lambda M^3}{r^7} + \mathcal{O}\left(\frac{M^3 \lambda^2}{r^{11}} \right),
\end{align}
and clearly dominates over the super-exponentially decaying  homogenous solution at large $r$, thereby giving 
\be 
f(r) \approx 1 - \frac{2 M}{r} + h_{\rm p} \, .
\ee

Neither the near horizon approximation nor the asymptotic solution is valid in the entire spacetime outside of the horizon.  One means to bridge this gap is to numerically solve the equations of motion in the intermediate regime.  The idea is quite simple: For a given choice of $M$ and $\lambda$, pick a value for the free parameter $a_2$.  Use these values in the near horizon expansion to obtain initial data for the differential equation just outside the horizon:  
\begin{align}\label{eqn:nh_data}
f(r_+ + \epsilon) &= 4 \pi T \epsilon + a_2 \epsilon^2 \, ,
\nn\\
f'(r_+ + \epsilon) &= 4 \pi T  + 2 a_2 \epsilon \, ,
\end{align}   
where $\epsilon$ is some small, positive quantity. A generic choice of $a_2$ will excite the exponentially growing mode in~\eqref{eqn:asymp_homog}. Thus, $a_2$ must be chosen extremely carefully and with high precision to obtain the asymptotically flat solution.   A satisfactory solution will be obtained if for some value of $r$ that is large (compared with the other scales in the problem), the numeric solution agrees with the asymptotic expansion to a high degree of precision.  In practice, we find that there is a unique value of $a_2$ for which this occurs. Of course, since the differential equation is very stiff, the numerical scheme will ultimately fail at some radius, $r_{\rm max}$.  The point at which this failure occurs can be pushed to larger distance by choosing $a_2$ more precisely and increasing the working precision, but this comes at the cost of increased computation time.\footnote{A solution for $r < r_+$ can be obtained by choosing $\epsilon$ to be small and negative in~\eqref{eqn:nh_data}.  The numerical scheme encounters no issues in this case. }   

\begin{figure*}[htp]
\centering
\includegraphics[width=0.32\textwidth]{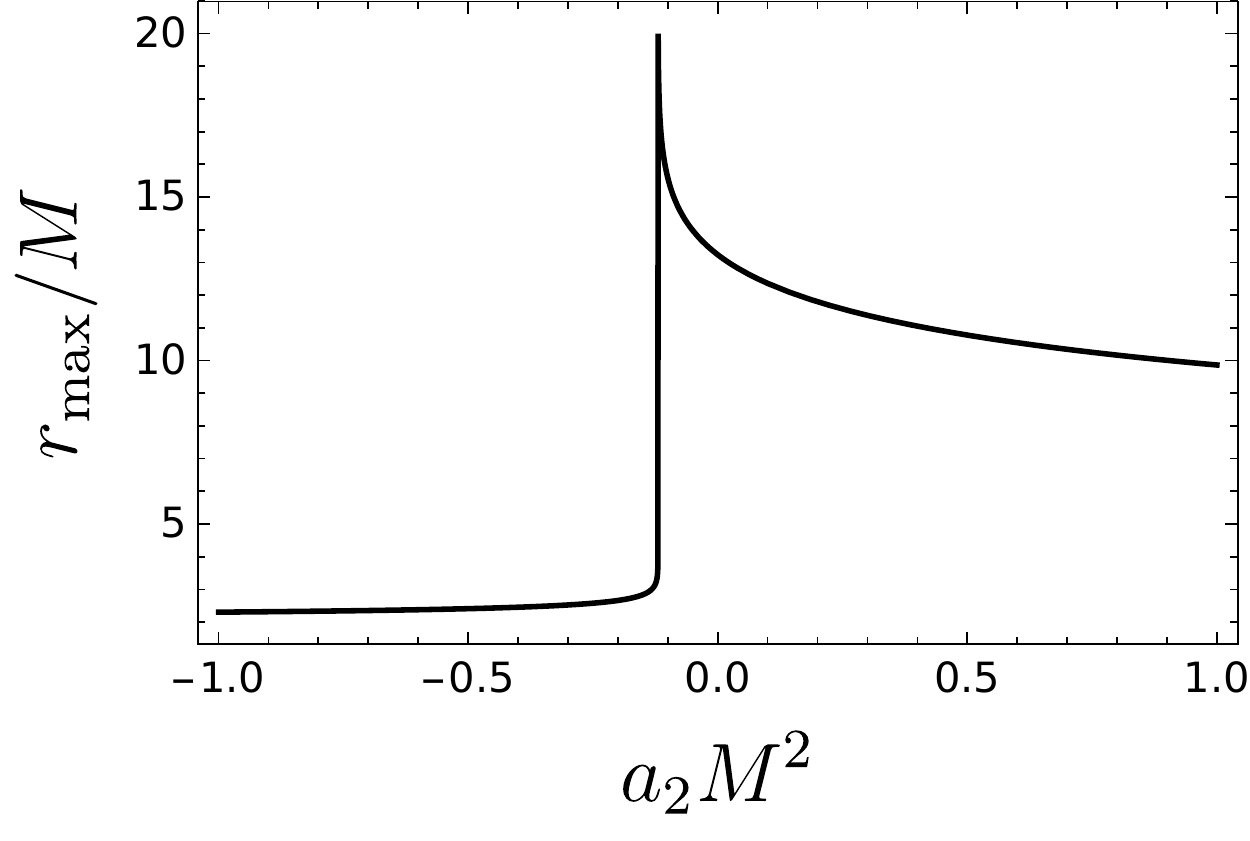}
\includegraphics[width=0.32\textwidth]{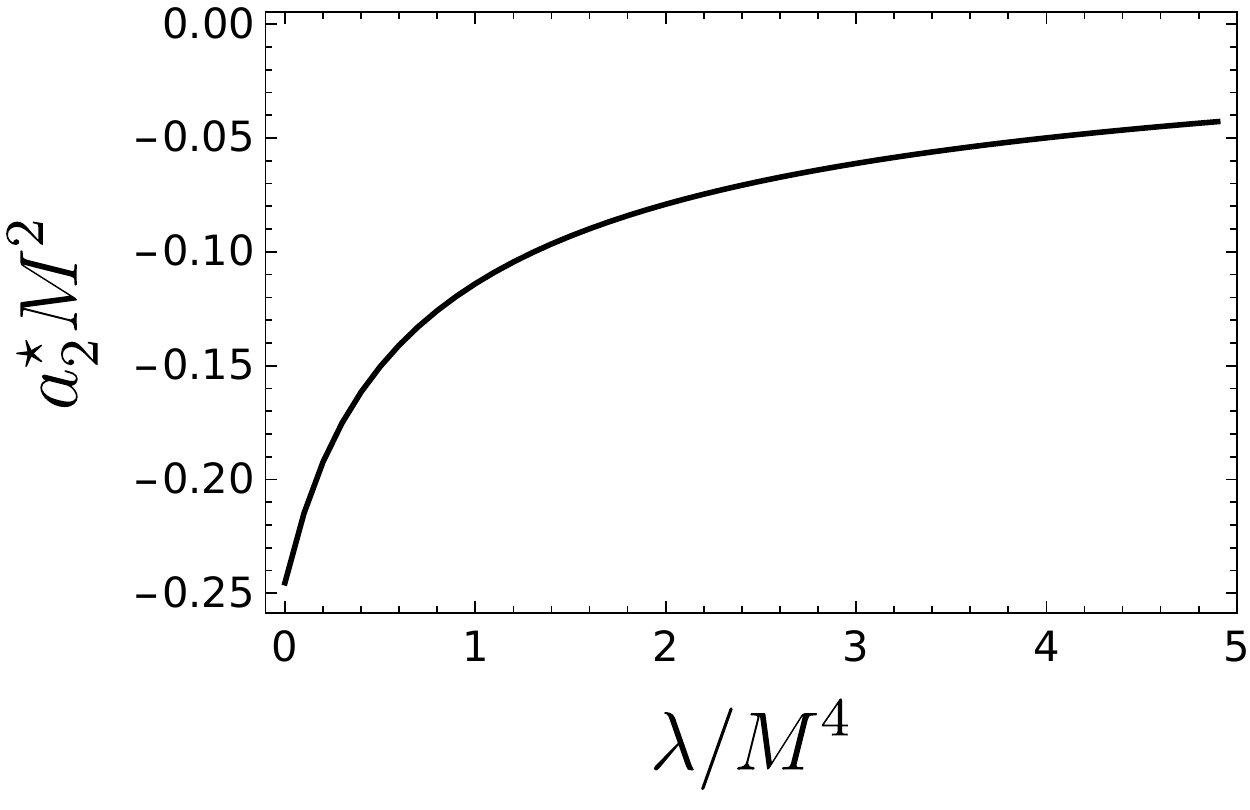}
\includegraphics[width=0.32\textwidth]{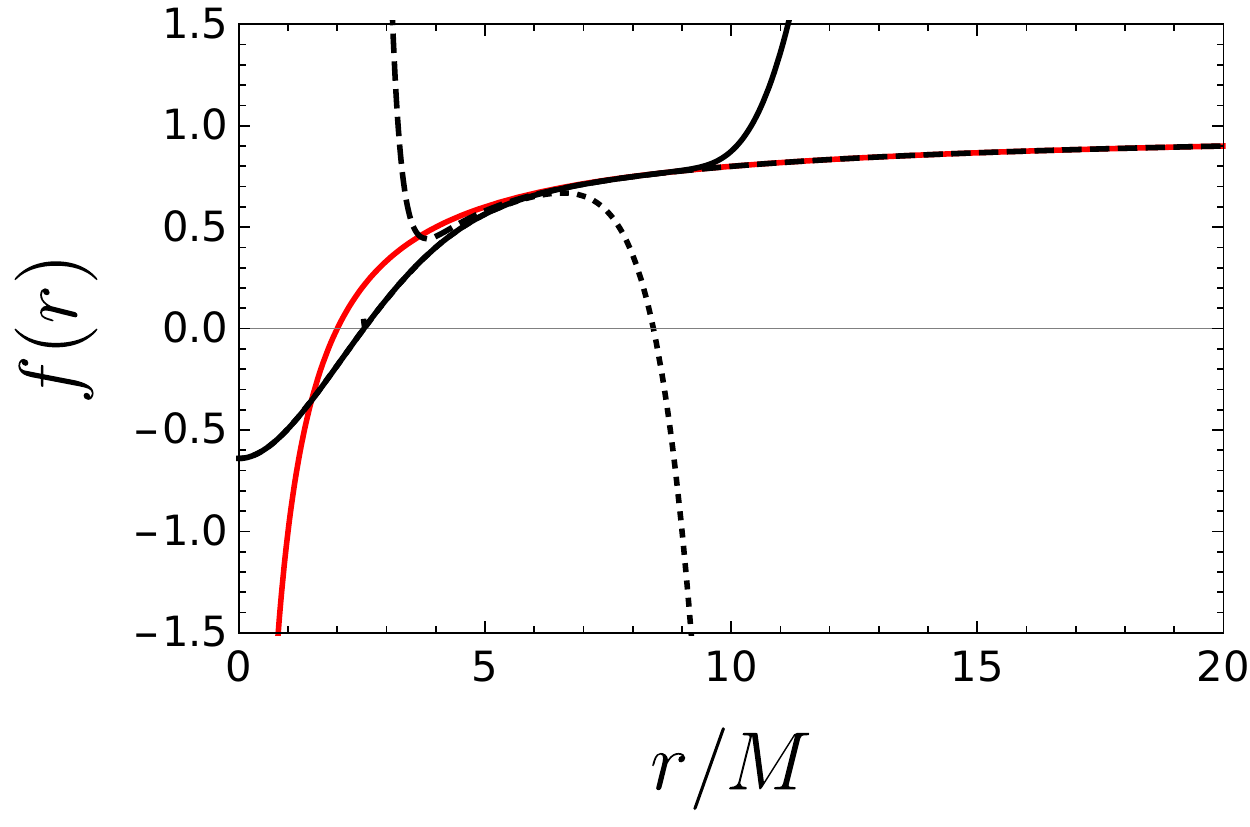}
\caption{{\bf Numerical scheme}: \textit{Left}: A plot of $r_{\rm max}$ (where the numerical solution breaks down) vs. $a_2$ for the case $\lambda = 1$. The peak corresponds to the value of $a_2$ that gives an asymptotically flat solution.  \textit{Center}: A plot of  the value of $a_2$ giving an asymptotically flat solution vs. $\lambda$.  Note that in the limit $\lambda \to 0$ we have $a_2 M^2 \to -1/4$, which coincides with the Einstein gravity result.  \textit{Right}: Numerical solution for $\lambda/M^4 = 10$ and a shooting parameter $a_2^\star = -0.022853992336918507$.  
The solid, red curve is the Schwarzschild solution of Einstein gravity.  The black, dotted curve is the near horizon approximation, including terms up to order $(r-r_+)^8$.  The dashed, black curve is the asymptotic solution, including terms up to order $r^{-12}$.  The solid black line is the numeric solution. In all cases, $\epsilon=10^{-6}$ was used in Eq.~\eqref{eqn:nh_data} to obtain the initial data.}
\label{fig:numerical_scheme}
\end{figure*}

In Fig.~\ref{fig:numerical_scheme} we highlight some sample numerical results.  The leftmost plot displays $r_{\rm max}$ vs. $a_2$, revealing a prominent peak at a point $a_2^\star$. The peak coincides with the value of $a_2$ that produces the asymptotically flat solution. In the center plot, we show the value of $a_2^\star$ plotted against the coupling, $\lambda$.  Notably, $a_2^\star$ limits to the Schwarzschild value of $a_2^\star M^2 = -1/4$ when $\lambda \to 0$.  While we have not been able to deduce a functional form for $a_2^\star$ from first principles,\footnote{See Appendix~\ref{app:shooting} for progress in this direction.} it is possible to perform a fit of the numeric results giving
\be\label{eqn:approx_a2} 
a_2^\star \left(x= \lambda / M^4 \right) \approx -\frac{1}{M^2} \frac{1 + 2.1347 x + 0.0109172 x^2}{4 + 15.5284 x + 8.03479 x^2},
\ee
which is accurate to three decimal places or better on the interval $\lambda/M^4 \in [0, 5]$. 

In the rightmost plot of Fig.~\ref{fig:numerical_scheme} we show a numerical solution for $\lambda/M^4 = 10$
and compare it with the Schwarzshild solution, as well as the near horizon and asymptotic approximate solutions. For the same physical mass, the ECG black hole has a larger horizon radius than the Schwarzschild solution.  Note that the near horizon solution provides an accurate approximation from $r=0$ to about $r=5 M$, but then rapidly diverges to $f\to -\infty$. The numeric solution begins to rapidly converge to the asymptotic solution near $r = 4M$, but near $r = 10 M$ it breaks down: the stiff system causes the integrated solution to rapidly diverge to $f\to +\infty$.  This  is just a consequence of not choosing $a_2^\star$ to high enough precision in the numeric method, and the exponentially growing mode has been excited. Before the numeric solution breaks down, the asymptotic solution (dashed line)  is accurate to better than 1 part in 1,000 and so it can be used to continue the solution to infinity.   

While both the near horizon and asymptotic approximations are useful within their respective domains of validity, neither provides a good approximation of the solution everywhere outside of the horizon.  To obtain an approximate solution valid everywhere outside of the horizon, we employ a continued fraction approximation to the metric function. First, we compactify the spacetime interval outside of the horizon by working in terms of the coordinate
\be 
x = 1 - \frac{r_+}{r},
\ee
and then write the metric function as~\cite{Rezzolla:2014mua, Kokkotas:2017zwt}
\be\label{eqn:cfrac_ansatz} 
f(x) = x \left[1 - \varepsilon(1-x) + (b_0 - \varepsilon)(1-x)^2 + \tilde{B}(x)(1-x)^3 \right],
\ee
where
\be 
\tilde{B}(x) = \cfrac{b_1}{1+\cfrac{b_2 x}{1+\cfrac{b_3 x}{1+\cdots}}} \, .
\ee
Using the ansatz~\eqref{eqn:cfrac_ansatz} in the field equations at large $r$ $(x=1)$ one deduces that
\begin{align}
\varepsilon &= \frac{2 M}{r_+} - 1 \, ,
\nn\\
b_0 &= 0 \, .
\end{align} 
Next, expanding~\eqref{eqn:cfrac_ansatz} near the horizon ($x=0$), the remaining coefficients can be fixed in terms of $T$, $M$, $r_+$ and one free parameter, $b_2$.  We have
\be 
b_1 = 4 \pi r_+ T + \frac{4 M}{r_+} - 3,
\ee
whereas $b_2$ is related to the coefficient $a_2$ appearing in the near horizon expansion~\eqref{eqn:nh_expand} as
\be\label{eqn:frac_b2}
b_2 = - \frac{r_+^3 a_2 + 16 \pi r_+^2 T + 6(M-r_+)}{4 \pi r_+^2 T + 4 M - 3 r_+} \, .
\ee
All higher order coefficients are determined by the field equations in terms of $T$, $M$, $r_+$ and $b_2$ (or, equivalently, $a_2$). Their form rapidly becomes quite messy,   but they can be obtained easily using, e.g. \textrm{Mathematica}.  We present the general expressions for the next few terms in the appendix. Since $b_2$ is not fixed by the field equations its value must be manually input into the continued fraction.  The appropriate thing to do is to use the value of $a_2^\star$ (as determined through the numerical method) in Eq.~\eqref{eqn:frac_b2}.  While the numerical integration of the field equations is very sensitive to the precision with which $a_2^\star$ is specified, the continued fraction is much less so, and a good approximation is obtained even with just a few digits.

\begin{figure}[htp]
\centering
\includegraphics[width=0.45\textwidth]{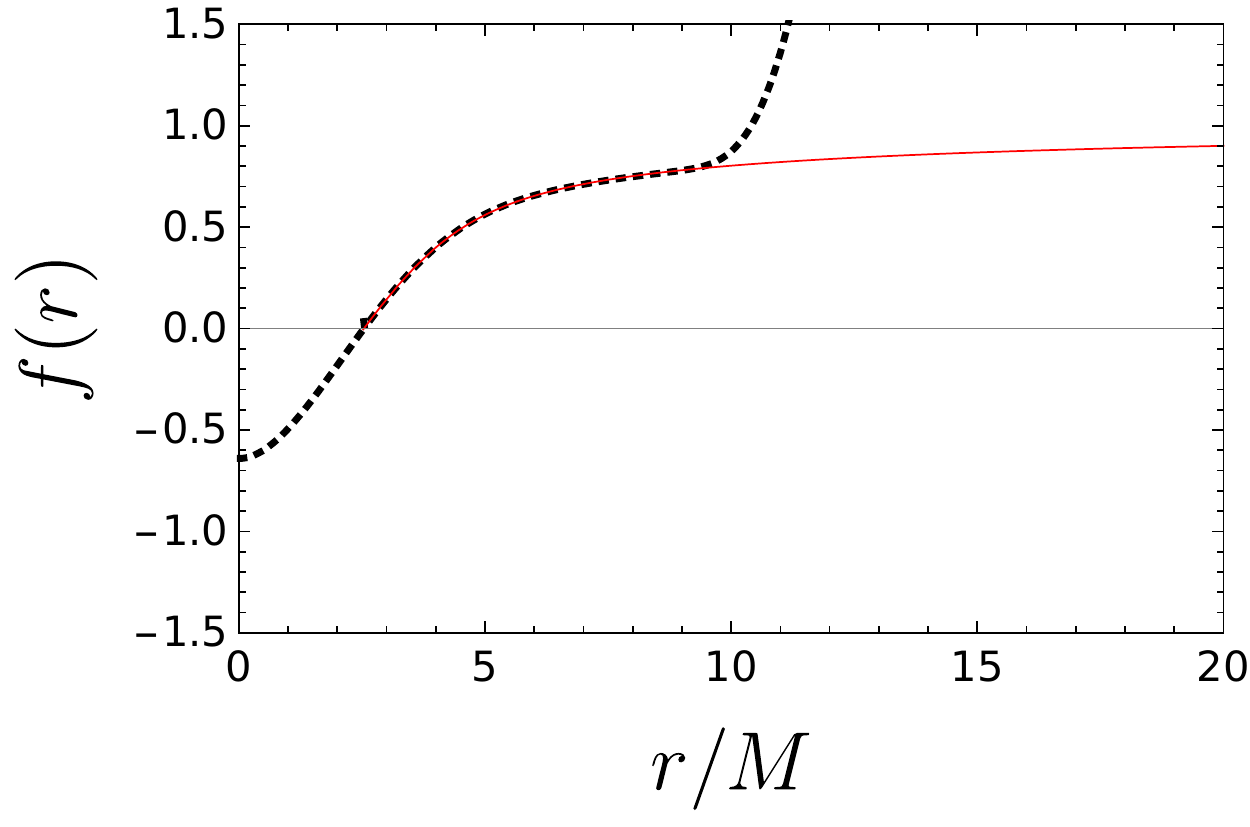}
\includegraphics[width=0.45\textwidth]{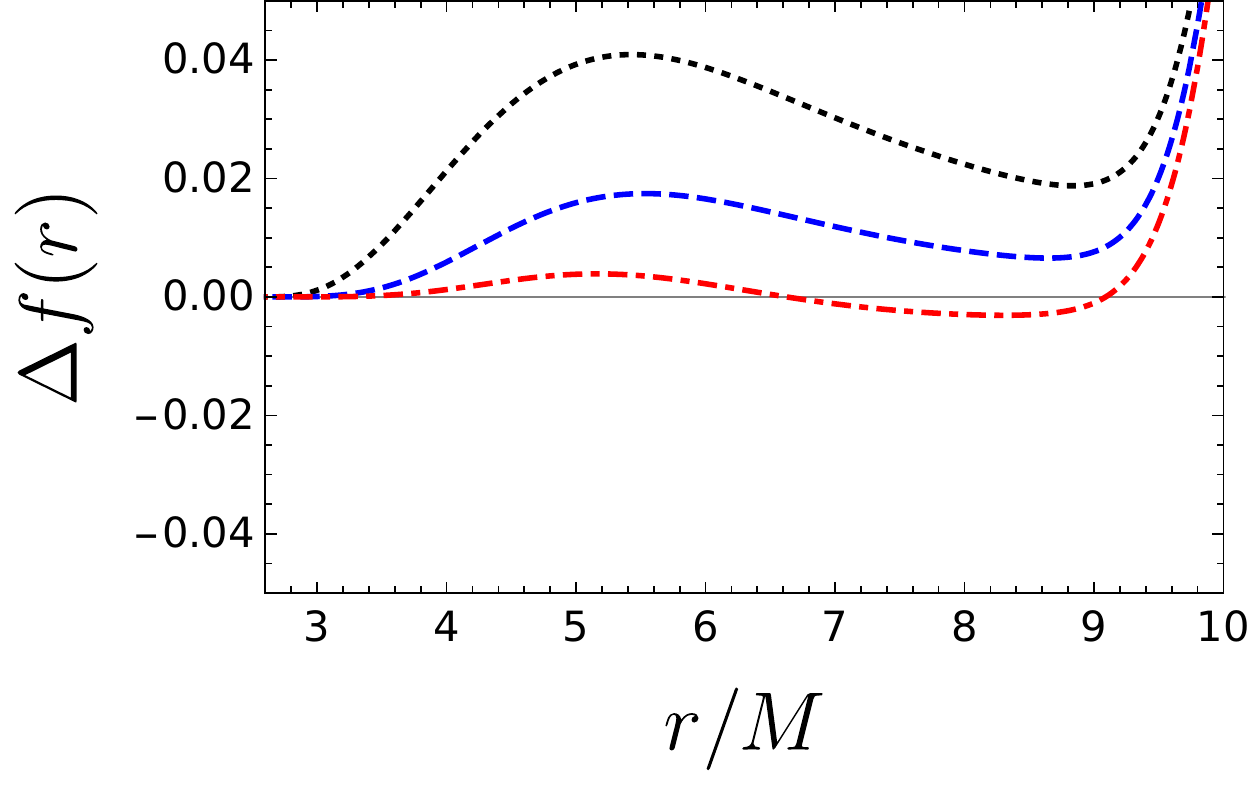}
\caption{\textbf{Continued fraction approximation}: \textit{Top}: Comparison of numeric solution (dotted, black) and continued fraction approximation (solid, red) for $\lambda/M^4 = 10$.
In the continued fraction, we have kept terms up to $b_5$; the continued fraction remains accurate even after the numeric solution fails. \textit{Bottom}: Difference between the metric function obtained numerically and via the continued fraction approximation keeping terms up to  $b_3$ (dotted, black), $b_4$ (dashed, blue), and $b_5$ (dot-dashed, red). }
\label{fig:cfrac_approx}
\end{figure}

Even at lowest order, the continued fraction approximation does a good job of approximating the solution everywhere outside the horizon. This only gets better as more terms are included.  In Fig.~\ref{fig:cfrac_approx} we display the continued fraction approximation when terms up to $b_5$ are retained.  We also show the difference between the numerical solution and the continued fraction approximation.  Where the numerical solution is valid, the continued fraction quite accurately approximates it. Furthermore, while the numerical solution fails at some sufficiently large distance, the analytic  approximation \eqref{eqn:cfrac_ansatz}
remains accurate everywhere outside the horizon.

\section{Properties of Black Hole Solutions}
\label{sec:BH_props}
 
\begin{figure}
\centering
\includegraphics[width=0.45\textwidth]{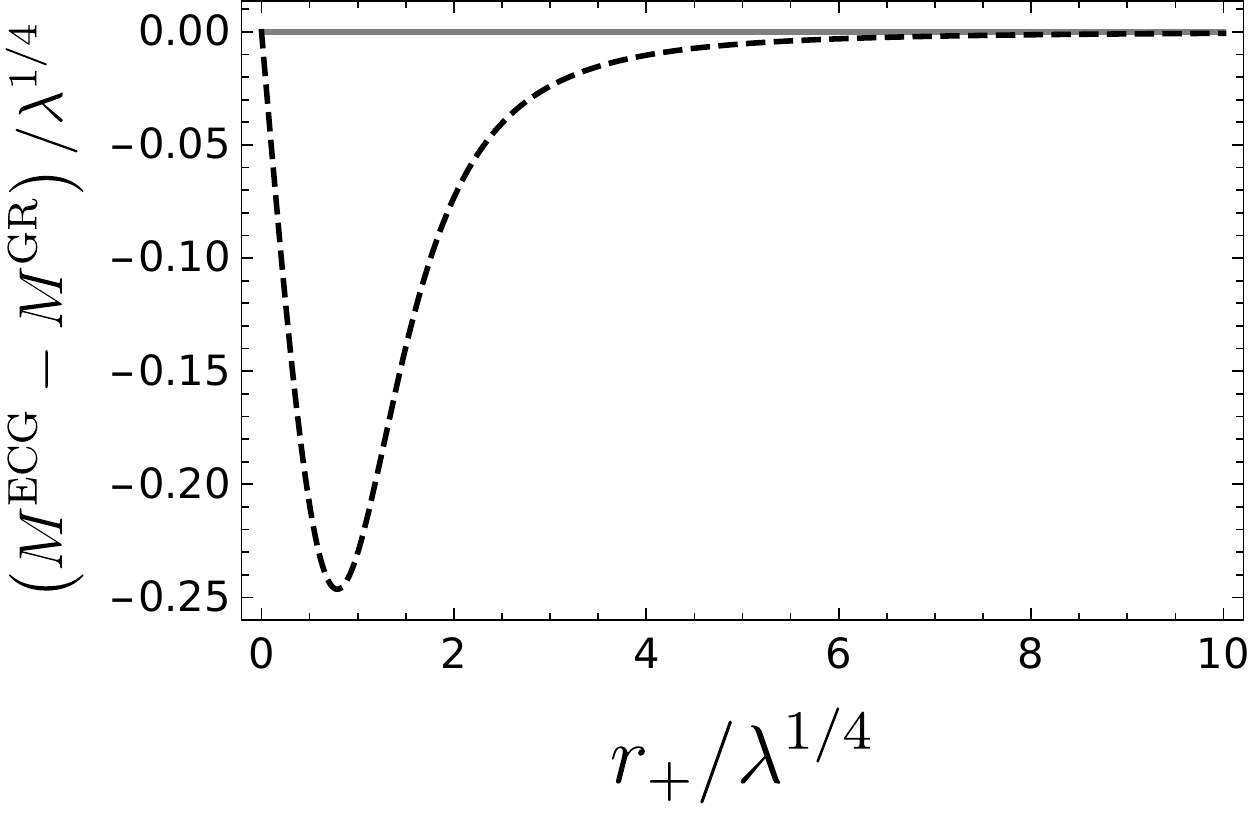}
\caption{{\bf Mass vs. horizon radius}: a plot of the difference between the ECG and Einstein black hole mass  vs. horizon radius. The plot illustrates a point of maximum difference.}
\label{fig:mass_diff}
\end{figure}

Let us now move on to consider some of the more interesting features of the black hole solutions in ECG. In the previous section we observed that, remarkably, despite the lack of an exact solution, the mass and temperature of these objects can be solved for exactly. A study of the black hole mass reveals that, for a given fixed $\lambda$, there exists a particular horizon radius for which the deviation from Einstein gravity is greatest. This is illustrated in Fig.~\ref{fig:mass_diff}.

Studying Fig.~\ref{fig:mass_diff}, we see that for both very small and very large black holes, the mass of the ECG black hole is very close to the mass of the ordinary Schwarzschild black hole. However, for intermediate values of horizon radius, there is a significant deviation. The horizon radius of maximum deviation can be solved for analytically yielding
\be 
r_+^{\rm dev} = \left( \frac{5 - \sqrt{\frac{11}{3}}}{8} \right)^{1/4} \lambda^{1/4} \ .
\ee
Substituting back into the expression for the mass yields approximately $M^{\rm dev} \approx 0.1476 \lambda^{1/4}$ for the mass of the black hole when the deviation from general relativity is maximal: a difference of about $37.5\%$ from the general relativity value. 

The ECG black holes depart from the Schwarzschild solution in another notable way: below a certain mass, the specific heat of these black holes becomes positive, indicating the onset of thermodynamic stability~\cite{Bueno:2017qce}. To see this, recall that the heat capacity is given by
\be 
C = \frac{\partial M}{\partial T} \ .
\ee
Using the expressions above for the mass and temperature, we find
\be 
C = \frac{2 r_+^2 \left(8 \lambda^2 - 6 r_+^4 \lambda + 3 \sqrt{r_+^4 + 4 \lambda} r_+^6 - 3 r_+^8 \right)}{\lambda \left(-r_+^2 + \sqrt{r_+^4 + 4 \lambda} \right)\left(-2 r_+^2 + \sqrt{r_+^4 + 4 \lambda} \right)},
\ee
from which a simple calculation reveals that the heat capacity is positive when the mass of the black hole satisfies,
\be 
M \le \frac{4}{3} \left(2 \sqrt{3} - 3 \right)^{3/4} \left(2 - \sqrt{3} \right) \lambda^{1/4} \, .
\ee

Next, we study the orbits of massive test bodies around the black hole in ECG. For such particles we have $g_{\alpha\beta}\dot{x}^{\alpha}\dot{x}^{\beta}=-\mu^2$, with $\mu$ denoting the rest mass of the infalling body.  Choosing coordinates so that  its orbit lies on the equatorial plane,  the geodesic motion is governed by the equation
\begin{equation}
\label{timelikegeo}
\dot{r}^2=\tilde{E}^2-f\left[1+\frac{\tilde{L}_{z}^{2}}{r^2}\right],
\end{equation}
where  $\tilde{E}$ and $\tilde{L}_{z}$ are the energy and angular momentum per unit rest mass $\mu$ of the body, with   a dot denoting the derivative with respect to proper time per unit rest mass~\cite{mtw}.

The second term on the right hand side of equation (\ref{timelikegeo}), 
\beq
\tilde{V}^2=f\left[1+\frac{\tilde{L}_{z}^{2}}{r^2}\right] \, ,
\eeq
acts like potential and we can investigate the motion on timelike geodesic from it by using the metric function obtained with the continued fraction method.

\begin{figure}[htp]
	\centering
	\includegraphics[width=0.45\textwidth]{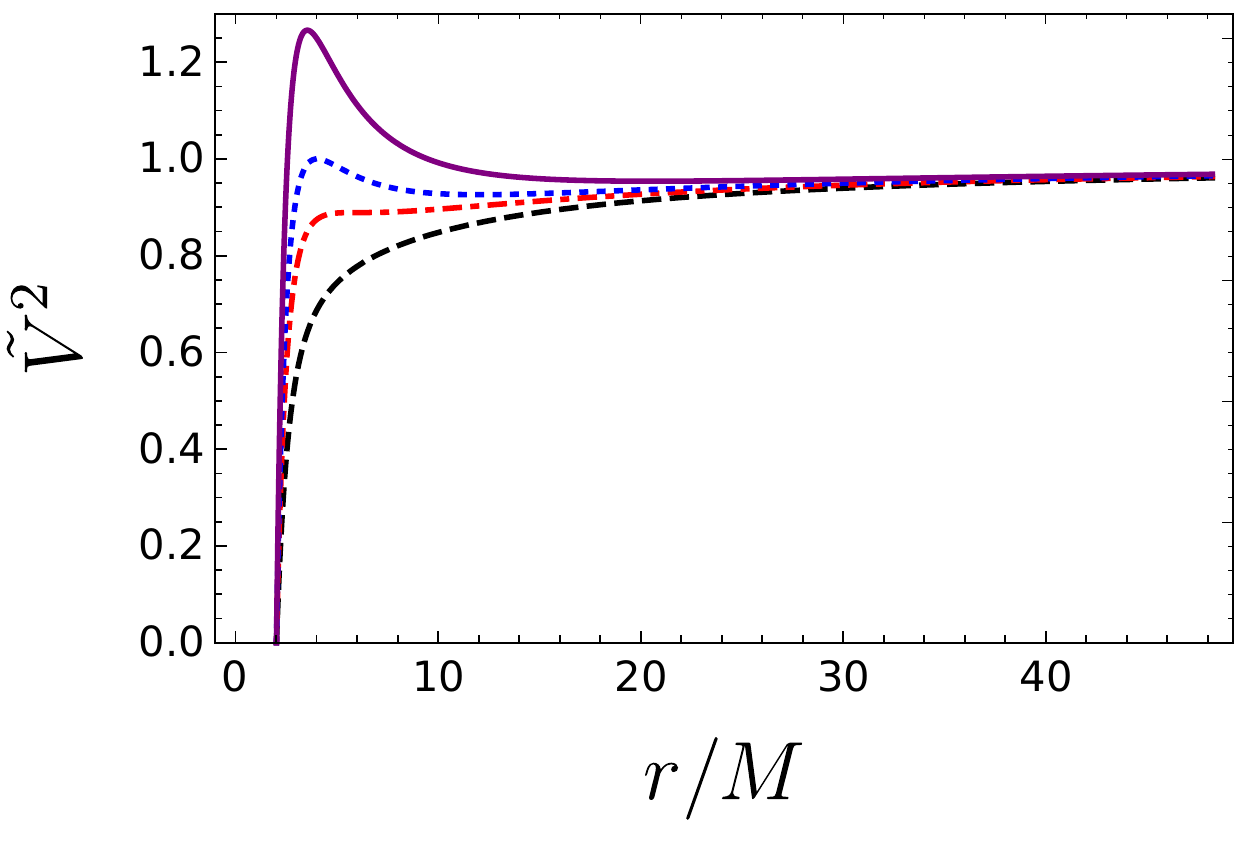}
	\caption{\textbf{Effective potential of infalling particle}: For $\lambda/M^4=0.1$ the effective potential is plotted for $\tilde{L}_{z}\approx 2.452 M$ (black, dashed curve), $\tilde{L}_{z}=\tilde{L}_{z, {\rm ISCO}}\approx 3.467 M$ (red, dot-dashed curve), $\tilde{L}_{z}\approx 4.005 M$ (blue, dotted curve), and $\tilde{L}_{z}\approx 4.903 M$ (purple, solid curve). The blue dotted curve with $\tilde{L}_{z}\approx 4.005 M$ has a maximum of $1$. For a particle coming from infinity, $\tilde{L}_{z}\approx 4.005 M$ is the minimum angular momentum it can have to avoid falling into the hole. The red, dot-dashed curve shows a point of inflection which is the innermost stable circular orbit.}
	\label{fig:isco_potential}
\end{figure}

In Fig.~\ref{fig:isco_potential} we plot $\tilde{V}^2$ for $\lambda/M^4=0.1$ for different values of $\tilde{L}_{z}$. For large value of $\tilde{L}_{z}$ there are two extrema in the curve of $\tilde{V}^2$, with the maximum (minimum) at the unstable (stable) orbits.
By decreasing the value of $\tilde{L}_{z}$, the radius of the unstable equilibrium orbit increases and the radius of stable equilibrium orbit decreases. The ISCO is at the inflection point of $\tilde{V}^2$; this is $r=r_{\rm ISCO}\approx 6.028 M$, which happens for particles with $\tilde{L}_{z}=\tilde{L}_{z, {\rm ISCO}}\approx 3.467 M$. The corresponding values in general relativity are $r_{\rm ISCO}= 6 M$ and $\tilde{L}_{z,ISCO}\approx 3.464 M$. Recall that any bodies coming from infinity can be bounded only if $\tilde{V}^2>1$ (or equivalently if $\tilde{L}_{z} \gtrsim 4.005 M$) since $\tilde{E}^2 \ge 1$.

To find out how $r_{\rm ISCO}$ and $\tilde{L}_{z, {\rm ISCO}}$ change with $\lambda$, we use small $\lambda$ approximation of the metric function
\begin{equation}\label{eqn:faap}
f_{app}(r,\lambda)=1-\frac{2M}{r}-\frac{1419 (r/M)^2-8362 r/M+10136}{12 (65-61 r/M) r^4}\lambda.
\end{equation}
The difference between this function and the one obtained by continued fraction up to $b_5$ is less than $1$ part in $10, 000$  at $r=r_{\rm ISCO}$ for $\lambda/M^4 <1$. By considering that $r_{\rm ISCO}$ is the inflection point of 
\beq
\tilde{V}_{app}^2=f_{app}(1+\frac{\tilde{L}_{z, {\rm ISCO}}^{2}}{r^2}),
\eeq
we could find $r_{\rm ISCO}$ and $\tilde{L}_{z, {\rm ISCO}}$ for different values of $\lambda$.

\begin{figure}[htp]
	\centering 
	\includegraphics[width=0.45\textwidth]{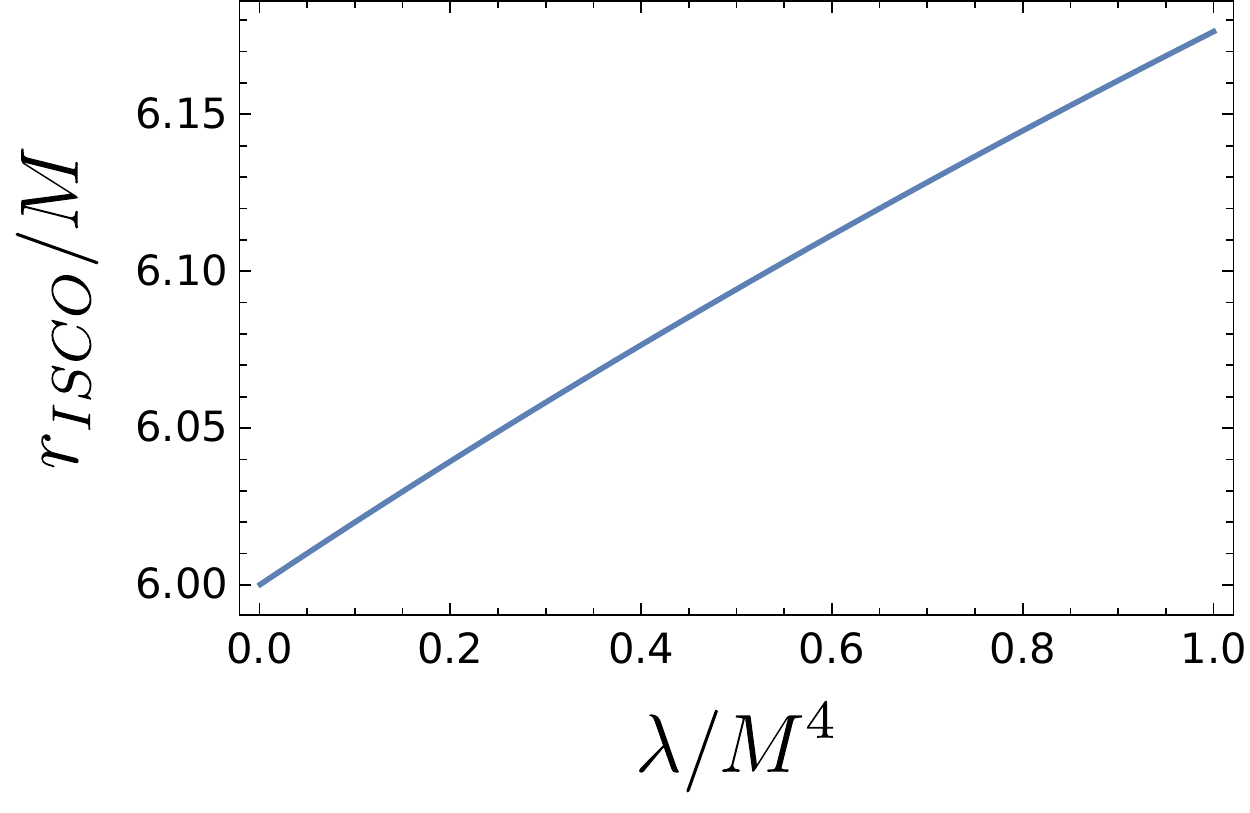}
	\includegraphics[width=0.48\textwidth]{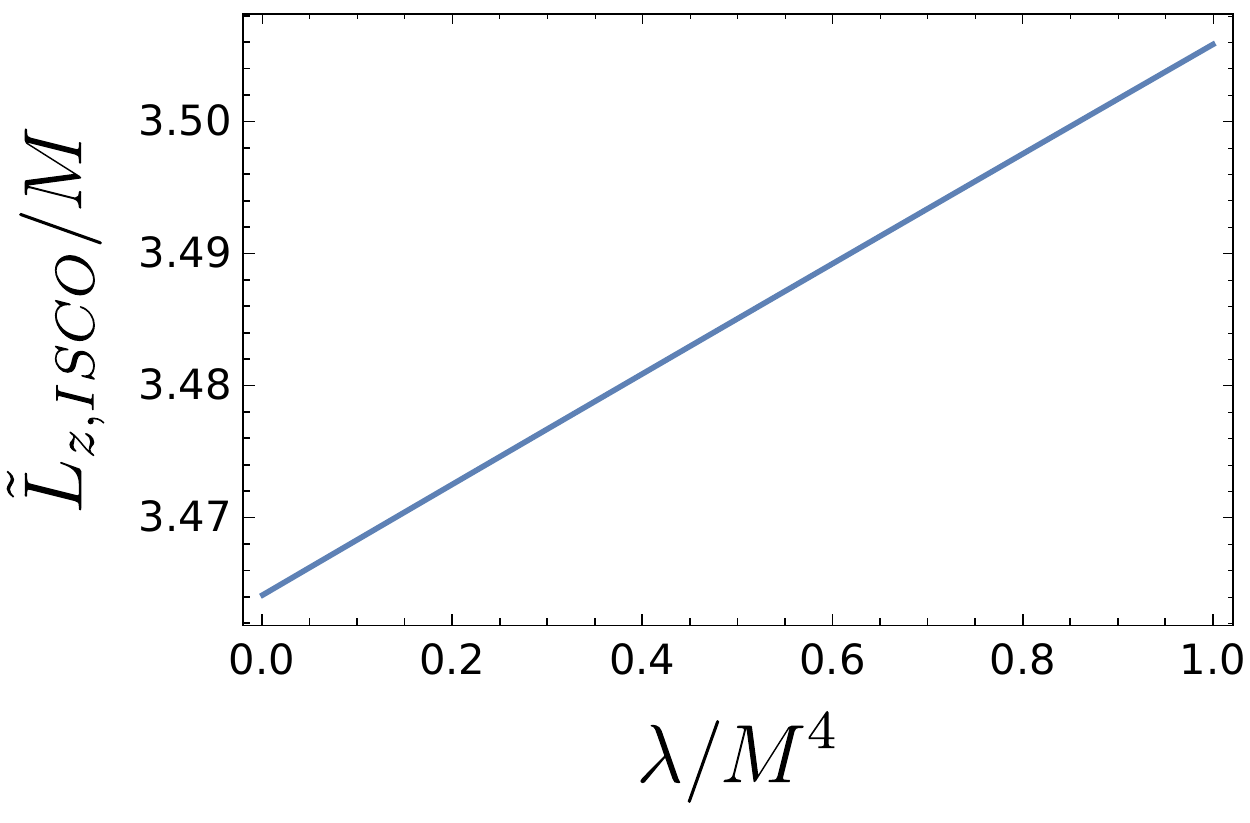}
	\caption{{\bf $r_{\rm ISCO}/M$ and $\tilde{L}_{z, {\rm ISCO}}/M$ vs. $\lambda/M^4$}: {\it Top}: Radius of the innermost stable circular orbit ($r_{\rm ISCO}$) as a function of the coupling constant of ECG. {\it Bottom}: The angular momentum for which the effective potential has an inflection point as a function of $\lambda$.}
	\label{fig:r_L_isco}
\end{figure}

In Fig.~\ref{fig:r_L_isco} we have plotted $r_{\rm ISCO}/M$ as a function of $\lambda/M^4$. By fitting the numerical results we find the relation
\begin{equation}
\label{apprisco}
r_{\rm ISCO}(\lambda)/M\approx\frac{6+0.00109641 \lambda/M^4}{1+0.00014898 \lambda/M^4},
\end{equation}
which is a small $\lambda$ approximation of $r_{\rm ISCO}$. Using (\ref{apprisco}) we obtain the approximate functional form
\begin{equation}
\frac{\tilde{L}_{z, {\rm ISCO}}}{M}\approx\sqrt{\frac{12+0.322149 \lambda/M^4}{1+0.0025491 \lambda/M^4}},
\end{equation}
 of the angular momentum at the ISCO, shown in the bottom plot of Fig.~\ref{fig:r_L_isco}. We can see that by increasing $\lambda$, both $r_{\rm ISCO}$ and the angular momentum of the orbiting particle at $r=r_{\rm ISCO}$ increases from their values in general relativity.

\section{Constraining ECG}
\label{sec:testECG}

From the analysis presented above it is clear that, for SSS solutions, the ECG corrections are most important when considering effects on scales near the horizon.  For distances a few times the horizon radius, the metric rapidly tends to the Schwarzschild solution.  In fact, one finds that the post-Newtonian parameter $\gamma$ is unity in ECG, just as in Einstein gravity.  This means that, in the weak field regime, the deviations caused by ECG will be small. To concretely illustrate just how small, here we will consider how the Shapiro time delay, the most stringent of the solar system tests, constrains ECG.

Here we will follow Weinberg's treatment~\cite{weinberg1972}. The time for a photon to travel between the points $r_0$ and $r$ is given by the integral
\be 
t(r, r_0) = \int_{r_0}^r \left[ f^2(r) \left(1 - \frac{f(r)}{f(r_0)} \left(\frac{r_0}{r} \right)^2 \right) \right]^{-1/2} dr \, .
\ee
For all practical purposes, the first few terms in the asymptotic expansion for $f(r)$ can be used when evaluating this integral. It is, of course, straightforward to perform the integration numerically and this is the method we employ.   However, it is illuminating to consider, via approximation, the first corrections to the Shapiro delay due to ECG analytically. Schematically the expression takes the form,
\be 
	t(r, r_0) = t^{\rm SR} (r,r_0)+ \Delta t^{\rm GR} (r,r_0) + \Delta t^{\rm ECG}(r, r_0) \, ,
	\ee
	where $t^{\rm SR} (r,r_0)=\sqrt{r^2 - r_0^2}$ is the contribution that would arise from light propagating in flat spacetime. The general relativity correction is well-known, and the first few terms take the form,
	\begin{align}
	\Delta t^{\rm GR} (r,r_0) =& 2 M \ln \left[ \frac{r + \sqrt{r^2 - r_0^2}}{r_0} \right]
	\nn\\
	&+ M\left(\frac{r-r_0}{r + r_0} \right)^{1/2} + \cdots \, .
	\end{align}
	The higher order corrections are easily computed, but become increasingly complicated and so we do not present them here. To lowest order in $M/r$, $M/r_0$ and $\lambda/M^4$, the ECG correction takes the form:  
	\begin{align}
	\Delta t^{\rm ECG}(r, r_0) =& 
	\nn\\
	\frac{ \lambda }{ M^4} \bigg[ \frac{189  \pi M^6}{r_0^5} &+  18 \sqrt{1 - \frac{r_0^2}{r^2}}\frac{M^6  \left(6 r_0^2 + 13 r^2 \right)}{r_0^4 r^3} 
	\nn\\
	&- \frac{378 M^6}{r_0^5}\arctan \left(\frac{r_0}{\sqrt{r^2-r_0^2}} \right)	
	\bigg]  \, .
	\end{align}
This expression shows how incredibly suppressed the corrections due to ECG are at the level of solar system tests: Taking the mass above to be a solar mass ($M_{\astrosun} = 1477 \, {\rm m}$ and taking the smallest possible value of $r$ as the radius of the sun, $r_{\astrosun} = 6.957 \times 10^{8} {\rm m}$, then the factor in square brackets is roughly $(M_{\astrosun}/r_{\astrosun})^5 \sim 10^{-29}$. Therefore $\lambda/M^4$ can actually be very large while maintaining agreement with solar system tests of general relativity.

For a radar signal traveling from Earth to Mercury, grazing the sun along the way, the time delay is,
\begin{align} 
(\Delta t)_{\rm max} =& 2 \bigg[ t(r_{\earth} , r_{\astrosun} ) + t(r_{\astrosun} , r_{\mercury} ) 
	\nn\\
	&- \sqrt{r_{\earth}^2 - r_{\astrosun}^2} -  \sqrt{ r_{\mercury}^2 - r_{\astrosun}^2} \bigg].
\end{align} 
Deviations of this result from the prediction of general relativity have been constrained to be less than $0.0012 \%$~\cite{Will:2014kxa}. A careful numerical evaluation of the integrals reveals that provided
\be 
\lambda < 4.57 \times 10^{22} M_{\astrosun}^4,
\ee 
ECG will be consistent with constraints arising from the Shapiro time delay experiment. The surprising size of this value reiterates that the deviations from general relativity are most significant in the vicinity of a black hole horizon.

\section{Black Hole Shadows}
\label{sec:shadow}

The key advantage of the continued fraction approximate solution is that it is valid everywhere outside the horizon. As such, it can be used in the same way that an analytic solution could be used. This opens the door to the study of a variety of interesting questions that would be considerably more difficult if we were able to use only the numerical solution. Here we use the approximate solution to explore black hole shadows in this theory.  In this section, for a generic spherically symmetric black hole, we present an equation for the angular radius of the black hole shadow as seen by a distant observer. Then, by using approximate analytic solution, we will demonstrate that the shadow of the black hole gets bigger as the coupling constant of ECG increases.

Consider the generic spherically symmetric line element
\begin{equation}
\label{lelem}
ds^{2}=-fdt^{2}+\frac{dr^{2}}{f}+r^{2}(d\theta^{2}+\sin^{2}\theta d\phi^{2}),
\end{equation}
and the Lagrangian
\begin{equation}
\label{lag}
\mathcal{L}=\frac{1}{2}g_{\mu\nu}\dot{x}^{\mu}\dot{x}^{\nu}=\frac{1}{2}(-f\dot{t}^{2}+\frac{\dot{r}^{2}}{f}+r^{2}\sin^{2}\theta\dot{\phi}^{2}).
\end{equation}
Suppose a light ray travels toward the black hole. We can always choose the coordinates so that it stays on the equatorial plane.  Its energy and angular momentum
\begin{equation}
\label{consquans}
E=-\frac{\partial \mathcal{L}}{\partial \dot{t}}=f\dot{t},
\qquad
L_{z}=\frac{\partial \mathcal{L}}{\partial \dot{\phi}}=r^{2}\dot{\phi},
\end{equation}
are, of course, constant since they are conserved. 

For null geodesics $\mathcal{L}=0$. After some straightforward calculations we can write equation (\ref{lag}) as \cite{synge}
\begin{equation}
\label{nulllag}
\left(\frac{dr}{d\phi}\right)^{2}=r^{4}\left(\frac{1}{\xi^{2}}-\frac{f}{r^{2}}\right),
\end{equation}
for $\theta=\pi/2$, in which we have conventionally used $\xi=L_{z}/E$ as the constant of motion.

Since the left hand side of equation (\ref{nulllag}) is non-negative, we have
\begin{equation}
\label{xicon}
\xi^{2}\leq\frac{r^{2}}{f}.
\end{equation}
If $\xi^{2}$ is less than the minimum of $r^{2}/f$, the coordinate $r$ of the light ray will be always decreasing and the light ray will eventually reach the horizon. For $\xi^{2}$ bigger than the minimum of $r^{2}/f$, the light ray will escape to infinity after getting to a minimum distance $r_{*}$ from the black hole defined by $\xi^{2}=r_{*}^{2}/f(r_{*})$.

The boundary between these two types of light rays is given by the minimum of $r^{2}/f$. This defines the photon sphere around a (spherically symmetric) black hole. Since the paths of outgoing and ingoing light rays are the same, we can think of the light rays reaching us from an arbitrarily large number of sources after deflecting from the black hole. The shadow of the black hole is in fact the shadow of the photon sphere.

Denoting  the inclination angle of the light ray from the radial direction by $\delta$,   we can write \cite{synge}
\begin{equation}
\label{cot}
\cot\delta =\frac{1}{\sqrt{f}r}\left(\frac{dr}{d\phi}\right),
\end{equation}
 using the metric (\ref{lelem}) (with $\theta=\pi/2$).  From equations (\ref{nulllag}) and (\ref{cot}), we have
\begin{equation}
\sin^{2}\delta=\xi^{2}\frac{f}{r^{2}}.
\end{equation}
Denoting the radius of the photon sphere by $r_{ps}$ we have
\begin{equation}
\label{anrad}
\delta=\sin^{-1}\left(\sqrt{\frac{r_{ps}^{2}}{f(r_{ps})}\frac{f(D)}{D^{2}}}\right),
\end{equation}
which is the angular radius of the shadow as seen by an observer at $D$.

\section{Observational Tests of ECG}
\label{sec:sgr}

The current EHT project \cite{eht} will study the black hole at the center of our Galaxy. 
Present-day observation indicates that this black hole, Sgr A*,
has a mass $M=6.25\times 10^9 \, {\rm m}$ and its distance is $D=2.57\times 10^{20} \, {\rm m}$~\cite{mnd}. In general relativity 
its horizon radius is $r_{+}=2M$, and the radius of its photon sphere is $r_{\rm ps}=3M$. Using  (\ref{anrad}) we obtain the known result $\delta=26.05~\mu as$. 

We can exploit the above results to provide an observational test of ECG.  Suppose ECG is the appropriate theory of gravity in our Galaxy.  For sufficiently small $\lambda$ it will pass all solar system tests, but for sufficiently large black hole masses (such as Sgr A*) its predictions will differ notably from that of general relativity.
 
\begin{figure}[htp]
	\centering 
	\includegraphics[width=0.45\textwidth]{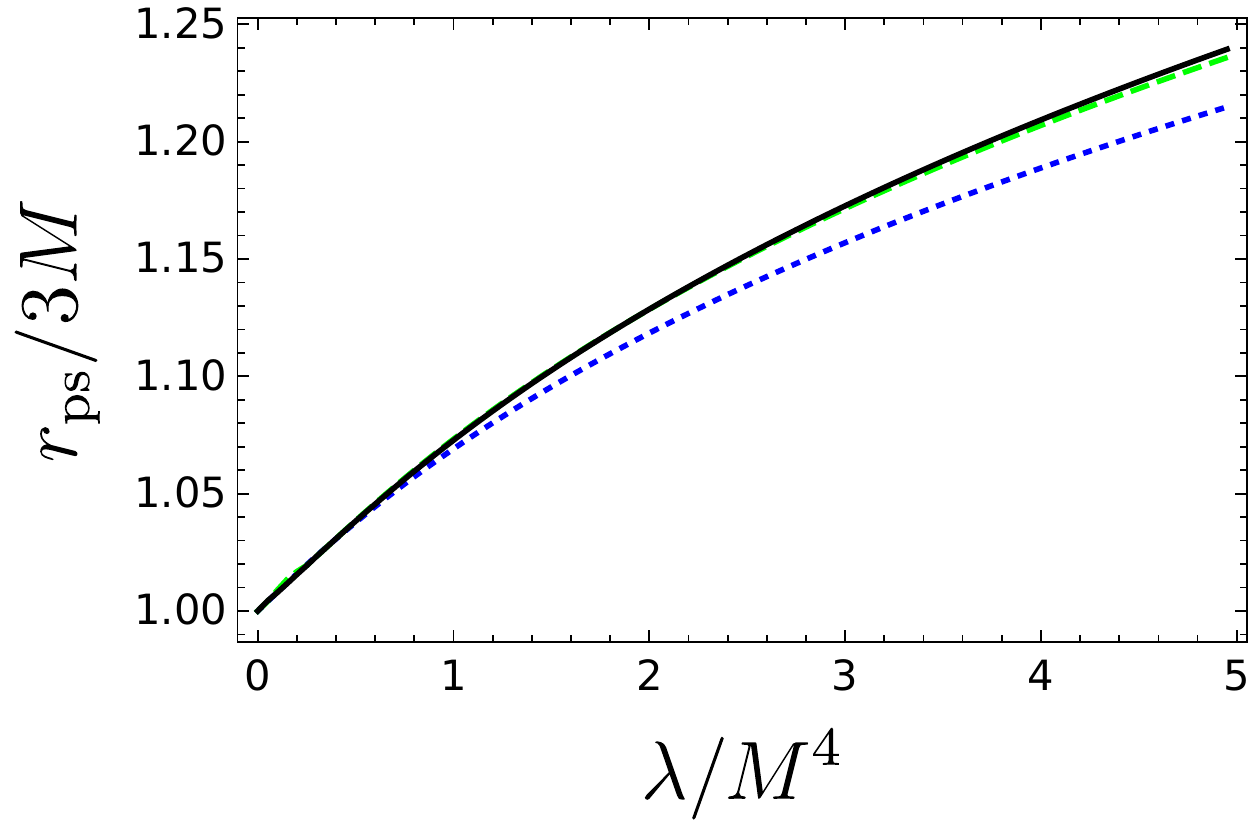}
	\includegraphics[width=0.48\textwidth]{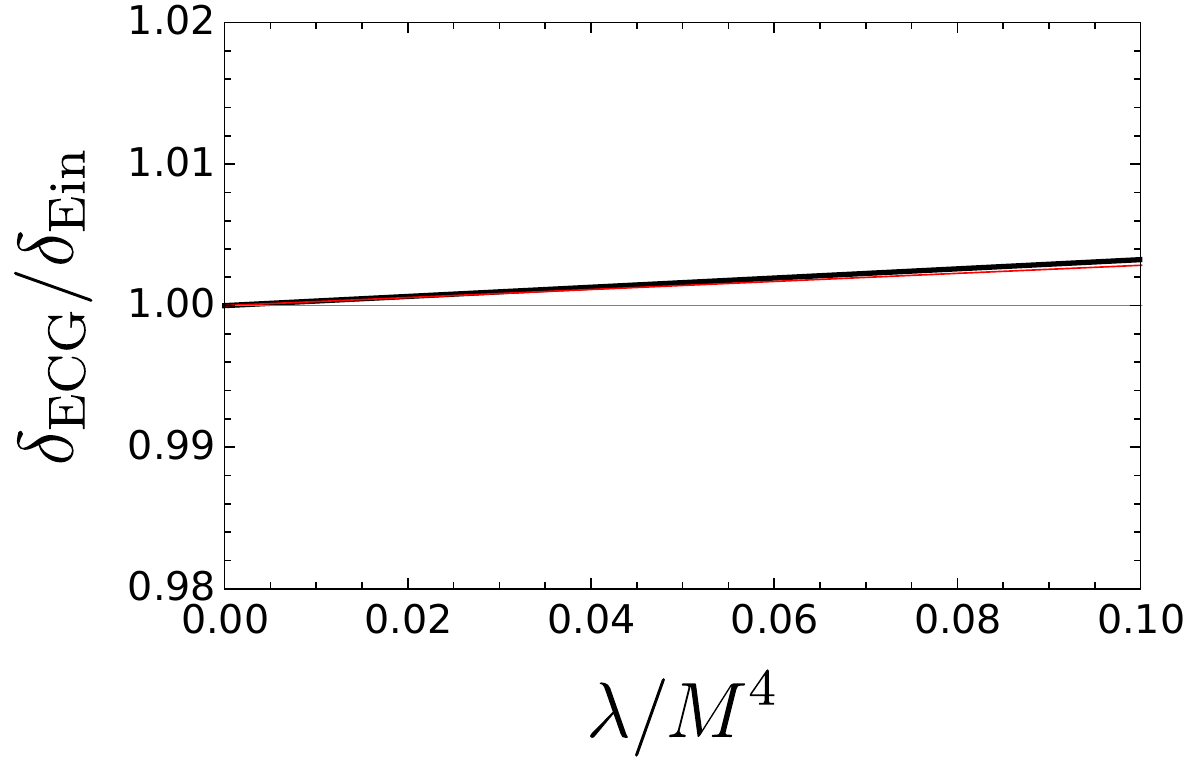}
	\caption{{\bf Photon sphere and angular radius}: {\it Top}: A plot of the photon sphere radius, $r_{\rm ps}$, vs. the ECG coupling computed using the continued fraction truncated at: $b_2$ (blue, dotted), $b_5$ (green, dashed) and $b_6$ (black, solid). For small coupling (compared to the mass), even the lowest order approximation is accurate, while for larger couplings the continued fraction converges after the inclusion of the first few terms (the black and green curves are virtually indistinguishable). The general result is that larger ECG coupling pushes the photon sphere to larger distances. {\it Bottom}: A plot of the ratio of the angular radius of the shadow for a black hole of mass $6.25 \times 10^{9} \, m$ and viewing distance $D = 2.57 \times 10^{20} \, m$. The solid black line is the result of the continued fraction, truncated with $b_5 = 0$.  The red line is a linear approximation for small coupling, shown in Eq.~\eqref{eqn:ang_rad_approx}.}
	\label{fig:rad}
\end{figure}

If ECG is correct, the metric outside of a spherically symmetric black hole will be given to excellent accuracy by the continued fraction approximation \eqref{eqn:cfrac_ansatz}, with $M$ and $D$ having the values given above for the case of Sgr A*.  The horizon radius in ECG will be larger than in general relativity, as can be determined from Eq. \eqref{eqn:mass_temp}. The photon radius is likewise enlarged as well, as we depict in Fig.~\ref{fig:rad}. It is a simple matter to compute the angular radius of the black hole shadow using~\eqref{anrad}. We present the results of this calculation, as determined through a continued fraction approximation truncated at $b_5$ in Fig.~\ref{fig:rad} for mass and distance choices relevant for Sgr A*.  We note that, when $\lambda/M^4$ is small, we can use Eq. (\ref{eqn:faap}) to find the following expansion for the angular radius of the shadow
\begin{equation}\label{eqn:ang_rad_approx} 
\delta_{\rm ECG} = \delta_{\rm Ein} + \frac{5.29015\times 10^{10}}{\sqrt{\frac{3 D^2}{f(D) M^2} - 81}} \frac{\lambda}{M^4} + {\cal O}(\lambda^2),
\end{equation}
(in $\mu as$) which appears in Fig.~\ref{fig:rad} as the thin, red curve and has a difference of less than one percent from the numerical results for any $\lambda \le 0.1$. ECG leads to larger black holes shadows than seen in Einstein gravity. Since, for larger distances, $f(D)$ is practically identical in both general relativity and ECG, the differences seen in Fig.~\ref{fig:rad} are a result of the modifications in the strong gravity regime near the horizon.  However, as one would expect from dimensional grounds, the modifications are relatively small for objects of large mass, requiring $\lambda/M^4 \approx 0.3$ before differences of 1\% occur.

\section{Conclusions}
\label{sec:con}

We have shown that continued fraction approximations can be used to accurately and efficiently approximate black hole solutions in Einsteinian cubic gravity. The approximations are valid everywhere outside of the horizon.  The key advantage is that the continued fraction  \eqref{eqn:cfrac_ansatz} can be used in place of an exact solution, allowing one to study problems that would be difficult to tackle if limited to only numerical solutions.  

We have taken a first step in this direction by employing the continued fraction approximation to study some interesting features of black holes in ECG. We have that, for a given value of mass, the ISCO for a massive test body, will be on a larger radius for a larger value of the coupling constant $\lambda$ of ECG. Also, the angular momentum of the body at the ISCO increases as $\lambda$ increases. Likewise, study of the lightlike geodesics reveals that ECG enlarges the shadow of the black hole.

As would be expected from dimensional analysis, the effect of ECG is relatively small unless the ratio $\lambda/M^4$ becomes larger, or the viewing distance $D$ becomes comparable to the radius of the photon sphere. For Sgr A*  we find that for the largest value of $\lambda$ allowed by Shapiro time delay, ECG enlarges the angular radius of the shadow by 5 parts per million.  Nowadays, the resolution of EHT's $1.3 mm$ groundbased very long baseline interferometry (VLBI) is a few tens of microarcseconds~\cite{Akiyama} which is about the shadow size of Sgr A* and M87. In this resolution the shadow predicted by general relativity and ECG are indistinguishable at least for static solutions.

 By increasing the maximum distance in a VLBI array, i.e. by adding some space stations, or observing at shorter wavelenghts,  EHT (or a similar project) might achieve finer resolutions in the future. However resolutions of better than 1 nanoarcsecond to observe the effects of ECG on Sgr A* shadow will be required.

A natural direction for future work would involve extending these results to compute shadows of rotating black holes in ECG. These are of more direct astrophysical relevance, and may present distinct angular-dependent features that could be observed.  Similar techniques as those presented here (see also~\cite{Younsi:2016azx}) could be used to obtain an approximate rotating black hole solutions in this theory.  

The continued fraction approach also offers the exciting possibility of addressing the linear stability of black hole solutions in this theory by simplifying the analysis of quasi-normal modes. Not only would this be of astrophysical relevance for the four dimensional models, but it would also be relevant in the context of holography for the asymptotically AdS solutions. We hope to address these and other questions in future work. 

\section*{Acknowledgements}
This work was supported in part by the Natural Sciences and Engineering Research Council of Canada.

\onecolumngrid

\appendix
\section{Explicit Terms in Continued Fraction}

Here we present additional terms that appear in the continued fraction expansion  \eqref{eqn:cfrac_ansatz}.
\begin{align}
b_3 &= \frac{1}{192 T r_+ \pi b_2 \lambda \left(\tfrac{1}{2} + \pi r_+ T \right)\left(M - \tfrac{3}{4} r_+ + \pi r_+^2 T \right) } \bigg[-4 \pi r_+^6 T (b_2 + 3) + (3 b_2 + 6) r_+^5 
\nn\\
&+ \left(-320 \lambda \pi^3 T^3  (b_2^2 + 7 b_2 + 16) - 4 M b_2 - 6 M\right) r_+^4 + 240 \pi^2 \lambda T^2 \left (b_2^2 + \frac{31}{5} (b_2 + 1)\right) r_+^3 
\nn\\
&- 448 T \pi \lambda \left(M T \left(b_2^2 + \frac{41}{7} b_2 + \frac{93}{14} \right) \pi - \frac{9}{28}b_2 - \frac{3}{4} \right)  r_+^2 + 96 \pi M T \lambda (b_2^2 + b_2 - 1) r_+ 
\nn\\
&- 128 \pi T M^2 \lambda \left(b_2 + \frac{3}{2} \right)^2   \bigg].
\end{align}

\begin{align}
b_4 =& -\frac{1}{192 \pi  b_2 b_3 \lambda  r_+^3 T \left(2 \pi  r_+ T+1\right) \left(4 M+r_+ \left(4 \pi  r_+ T-3\right)\right)}
\nn\\
&\times 32 \left(2 b_2+3\right)^3 \lambda  M^3+48 \left(2 b_2+3\right) \lambda  M^2 r_+ \left(2 \pi  \left(18 b_2^2+2 \left(7 b_3+50\right) b_2+113\right) r_+ T-4 b_2^2+\left(2 b_3-9\right) b_2-4\right)
\nn\\
&+12 M r_+^2 \left(-2 \left(-6 b_2^3+6 \left(2 b_3-1\right) b_2^2+\left(21 b_3+40\right) b_2+54\right) \lambda \right.
\nn\\
&\left.+16 \pi ^2 \left(40 b_2^3+\left(44 b_3+331\right) b_2^2+\left(8 b_3^2+137 b_3+994\right) b_2+898\right) \lambda  r_+^2 T^2 \right.
\nn\\
&\left.-4 \pi  \left(84 b_2^3+\left(44 b_3+615\right) b_2^2+\left(-16 b_3^2+15 b_3+1284\right) b_2+815\right) \lambda  r_+ T \right.
\nn\\
&\left.+\left(b_2^2+\left(b_3+4\right) b_2+4\right) r_+^4\right)+r_+^3 \left(36 \left(b_2+2\right) \left(3 b_2 \left(b_3+2\right)+14\right) \lambda \right.
\nn\\
&\left.+128 \pi ^3 \left(35 b_2^3+45 \left(b_3+7\right) b_2^2+6 \left(2 b_3^2+29 b_3+185\right) b_2+1658\right) \lambda  r_+^3 T^3\right.
\nn\\
&\left.-48 \pi ^2 \left(100 b_2^3+\left(96 b_3+870\right) b_2^2+\left(8 b_3^2+306 b_3+2734\right) b_2+2663\right) \lambda  r_+^2 T^2\right.
\nn\\
&\left.+24 \pi  \left(45 b_2^3+3 \left(b_3+116\right) b_2^2+\left(-24 b_3^2-75 b_3+719\right) b_2+360\right) \lambda  r_+ T
\right.\nn\\
&\left.+12 \pi  \left(b_2^2+\left(b_3+4\right) b_2+6\right) r_+^5 T-3 \left(3 b_2^2+3 \left(b_3+4\right) b_2+14\right) r_+^4\right).
\end{align}

\section{Analytical Derivation of the Shooting Parameter}
\label{app:shooting}

As mentioned in the main body of the text, it is in fact possible to derive, analytically, the form of the shooting parameter $a_2^\star$ by demanding a consistent Einstein gravity limit for the near horizon expansion.  Here we shall describe this process in more detail.  

Recall that, near the horizon, the metric function is expanded as
\be 
f_{\rm nh} (r) = 4 \pi T (r-r_+) + a_2^\star (r-r_+)^2 + \sum_{i=3} a_i(a_2^\star) (r-r_+)^i,
\ee
where the constants $a_n$ with $n > 2$ are determined by the field equations in terms of the parameter $a_2^\star$ and $M$, $T$ and $r_+$.  We will demand that this expansion has a smooth $\lambda \to 0$ limit. It turns out that this constraint is also enough to ensure that the near horizon expansion limits to that for the Schwarzschild solution
\be\label{eqn:schw_nh} 
f^{\rm Ein}_{\rm nh} = \sum_{i=1} (-1)^{i-1} \frac{(r-r_+)^i}{r_+^i} \, .
\ee

We proceed by writing
\be 
a_2^\star = g(\lambda),
\ee
and expand each of $a_n(a_2^\star)$ to lowest order in $\lambda$. For example, the expansion for the first two terms is
\begin{align} 
a_3(a_2^\star) &= \frac{g(0) r_+^3+r_+}{9 \lambda } + \frac{3 r_+^6 g'(0)-6 g(0)^2 r_+^4+34 g(0) r_+^2-14}{27 r_+^3} + {\cal O}(\lambda) \, ,
\nn\\
a_4(a_2^\star) &= +\frac{g(0) r_+^6+r_+^4}{216 \lambda ^2} + \frac{3 r_+^6 g'(0)-60 g(0)^2 r_+^4+89 g(0) r_+^2+68}{648 \lambda }
\nn\\
&+\frac{3 r_+^{10} g''(0)-240 g(0) r_+^8 g'(0)+2 r_+^6 \left(89 g'(0)+72 g(0)^3\right)-968 g(0)^2 r_+^4+1040 g(0) r_+^2-278}{1296 r_+^4} + {\cal O}(\lambda).
\end{align}
Clearly, for $a_3$ to have a smooth $\lambda \to 0$ limit, we must take
\be 
g(0) = - \frac{1}{r_+^2},
\ee
which also cures the $\lambda^{-2}$ divergence in $a_4$.  Then, for $a_4$ to have a smooth $\lambda \to 0$ limit, we must take
\be 
g'(0) = \frac{27}{r_+^6} \, .
\ee 
Interestingly, this choice for $g'(0)$ also ensures that
\be 
a_3(a_2^\star) = \frac{1}{r_+^3} + {\cal O}(\lambda),
\ee
which is precisely the value expected from the Schwarzschild solution.  This procedure continues in the obvious way: The expansion of $a_n$ for small $\lambda$ fixes $g^{(n-3)}(0)$, which in turn guarantees that the term $a_{n-1}$ limits to the Schwarzschild value from~\eqref{eqn:schw_nh}.

It is straight-forward, but computationally costly, to do this to arbitrary order.  We have computed $g^{(n)}(0)$ up to $n = 15$, finding:
\begin{align}\label{eqn:a2_derivatives}
g(0) &=  -\frac{1}{r_+^2},\quad g'(0)= \frac{27}{r_+^6},\quad g''(0)= -\frac{3384}{r_+^{10}},\quad g^{(3)}(0)= \frac{1320534}{r_+^{14}},\quad g^{(4)}(0)= -\frac{1151833248}{r_+^{18}},\quad
\nn\\
g^{(5)}(0) &= \frac{1875967406160}{r_+^{22}},\quad g^{(6)}(0)= -\frac{5107532147380800}{r_+^{26}},\quad g^{(7)}(0)= \frac{21544624968666695280}{r_+^{30}},\quad 
\nn\\
g^{(8)}(0) &= -\frac{133135416924677418585600}{r_+^{34}},\quad g^{(9)}(0)= \frac{1154324990320626883159054080}{r_+^{38}},\quad 
\nn\\
g^{(10)}(0)&= -\frac{13568049825205878205542081792000}{r_+^{42}},\quad g^{(11)}(0)= \frac{210227289858470130670513367566041600}{r_+^{46}},\quad
\nn\\
g^{(12)}(0) &= -\frac{4194920428540096167815139429105212006400}{r_+^{50}},\quad 
\nn\\
g^{(13)}(0) &= \frac{105700177837430847101072792547386798551142400}{r_+^{54}} ,
\nn\\
g^{(14)}(0) &= - \frac{3306987976911675043248786217918581692121564979200}{r_+^{58}} ,
\nn\\
g^{(15)}(0) &= \frac{126609498143560198473638841716966388468374445902592000}{r_+^{62}}\, .
\end{align}

\begin{figure}
\centering
\includegraphics[width=0.5\textwidth]{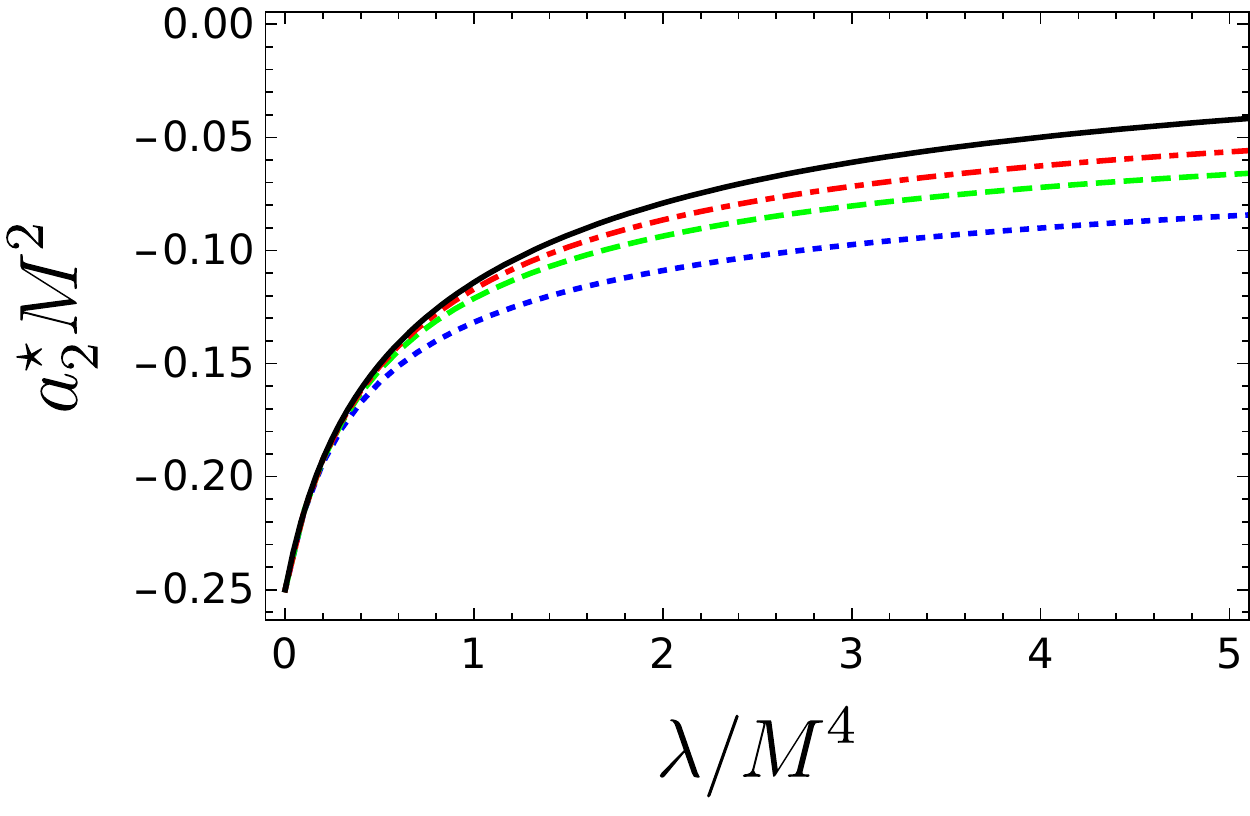}
\caption{{\bf Analytical approach for shooting parameter}: Here the solid black line denotes the value of $a_2^\star$ as determined through the numerical scheme.  The remaining curves denote Pad\'e approximants built from the derivatives presented in Eq.~\eqref{eqn:a2_derivatives}.  Specifically, the dotted, blue curve corresponds to a $2^{nd}$-order Pad\'e approximant, the dashed, green curve corresponds to a $4^{th}$ order Pad\'e approximant and the dot-dashed, red curve corresponds to a $7^{th}$ order Pad\'e approximant.  For $\lambda/M^4 < 1$ convergence to the numerical result is rapid, but convergence for larger values would require more derivatives than we were able to reasonably compute.}
\label{fig:pade_a2}
\end{figure}

While $g^{(n)}(0) \propto 1/r_+^{4n+2}$, we were not able to deduce the dependence of the coefficients of  $g^{(n)}(0)$ on $n$. 
The fact that these coefficents   grow unboundedly indicates that a Taylor series expansion of $a_2^\star = g(\lambda)$ has a small (perhaps vanishing) radius of convergence.  However, rather than a Taylor series we can use a Pad\'e approximant to reconstruct the form of $g(\lambda)$, and we show this in Fig.~\ref{fig:pade_a2}.\footnote{The Pad\'e approximant reveals why the Taylor series has unbounded coefficients: there is a simple pole located at small, negative $\lambda$ that a Taylor series approximation cannot capture.}  The basic conclusion is that, as more terms are included in the Pad\'e approximant, the form of $g(\lambda)$ converges to the results of our numerical scheme presented in Fig.~\ref{fig:numerical_scheme}.  While the convergence is fast for small $\lambda$, more terms are required to obtain good convergence for larger $\lambda$.  Thus with the fifteen derivatives presented in~\eqref{eqn:a2_derivatives}, it is not possible to accurately match $a_2^\star$ over the full domain of $\lambda$, and the fit to the numerical data~\eqref{eqn:approx_a2} is more accurate for larger $\lambda$.  If the functional dependence of $g^{(n)}(0)$ on $n$ could be deduced, then this would allow $a_2^\star$ to be determined to arbitrary precision. 

\section{Truncation method}

Consider the metric function written in continued fraction form (\ref{eqn:cfrac_ansatz}). If we are going to keep the fractions to $b_{n}$, we do so by setting $b_{n+1}=0$. Since all $b_m$'s $(m>2)$ are functions of $b_2$, we propose that, given the coupling constant $\lambda$ and the mass $M$ (or horizon radius $r_+$), the equation $b_{n+1}=0$ can be solved to find an approximate value for $b_2$. This way we can find an approximate equation for the metric function.

This method for finding $b_2$, which we call the truncation method, has some advantages over the numerical method. First, the truncation method is a much easier way to find $b_2$. We can choose whatever value of $\lambda$ and $M$ and solve $b_{n+1}=0$ to find $b_2$. Also, in the case that we kept the fraction to $b_3$, we have been able to solve $b_4=0$ for a generic $\lambda$ and $r_+$ to find $b_2$. So we have found an approximate analytical solution of ECG which is valid for all $\lambda$.

To clarify the agreement between the continued fraction (as obtained via the truncation method) and the numerical solution we refer the reader to Fig.~\ref{fig:cfrac_truc}. The agreement is quite remarkable, and while not quite as good as that when $b_2$ is obtained numerically (c.f. Fig.~\ref{fig:cfrac_approx}), the difference is small. The primary drawback of the truncation method is that the equation  $b_m=0$ $(m>2)$ usually results in multiple real solutions of $b_2$;  the right one must be chosen so that the metric function is not singular outside the horizon.

\begin{figure}[htp]
\centering
\includegraphics[width=0.45\textwidth]{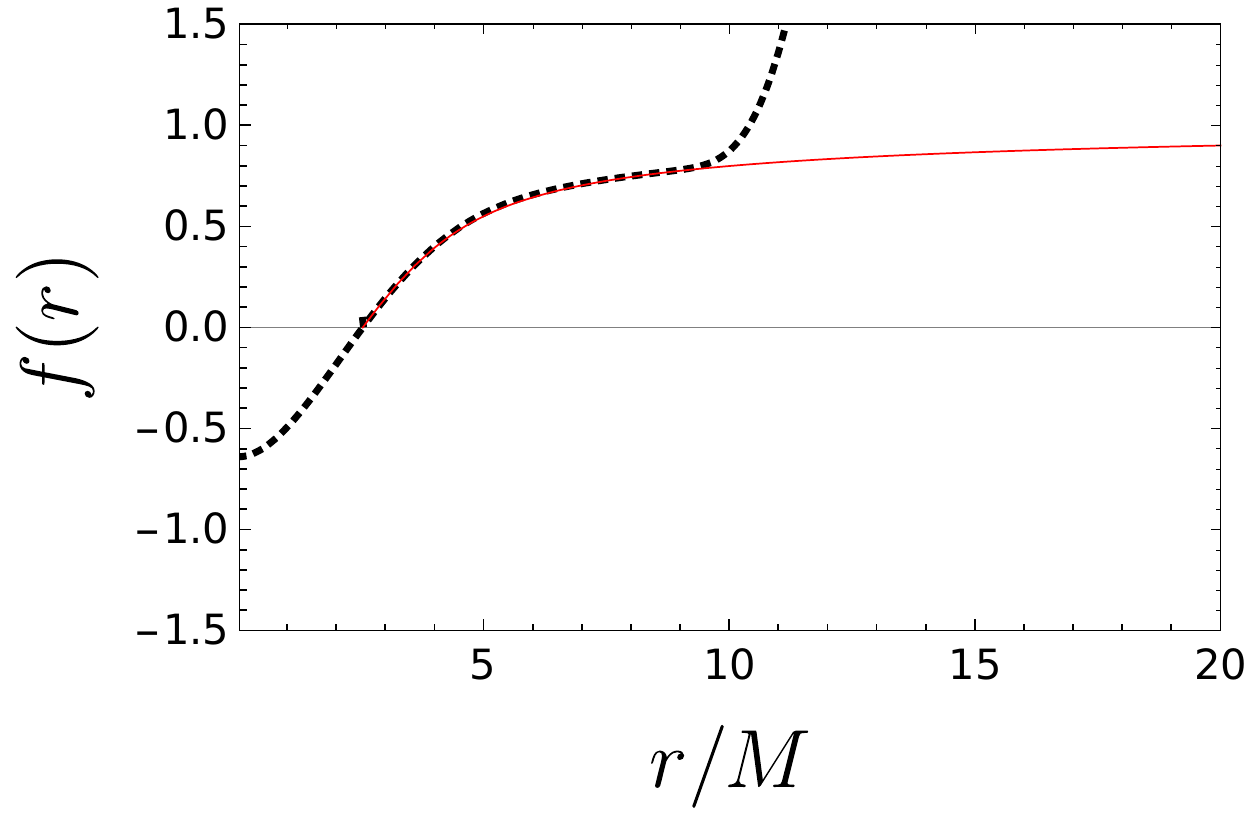}
\includegraphics[width=0.45\textwidth]{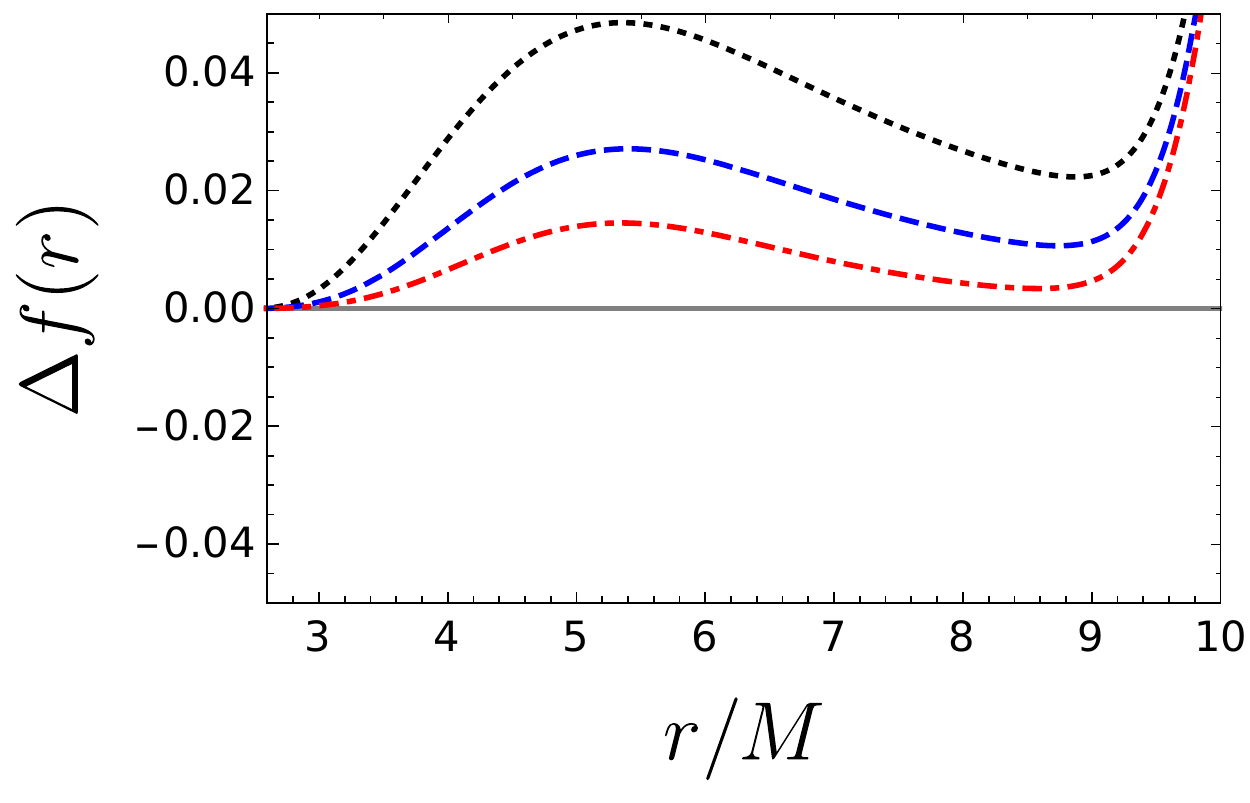}
\caption{\textbf{Continued fraction approximation with truncation method}: \textit{Left}: Comparison of numeric solution (dotted, black) and continued fraction approximation (solid, red) for $\lambda/M^4 = 10$.
In the continued fraction, we have kept terms up to $b_5$; here, we have found $b_2$ by solving $b_6=0$. \textit{Right}: Difference between the metric function obtained numerically and via the continued fraction approximation truncated after $b_3$ (dotted, black), $b_4$ (dashed, blue), and $b_5$ (dot-dashed, red).}
\label{fig:cfrac_truc}
\end{figure}

\twocolumngrid

\bibliography{mybib}
\end{document}